\documentclass[a4paper]{article}

\usepackage{INTERSPEECH2021}
\usepackage{todonotes,cite}
\usepackage{nidanfloat}
\usepackage{enumitem}
\usepackage{multirow}
\title{Flexi-Transducer: Optimizing Latency, Accuracy and Compute for\\Multi-Domain On-Device Scenarios}
\name{Jay Mahadeokar, Yangyang Shi\textsuperscript{*}\thanks{*Equal Contribution}, Yuan Shangguan\textsuperscript{*}, Chunyang Wu\textsuperscript{*}, Alex Xiao, Hang Su, Duc Le, Ozlem Kalinli, Christian Fuegen, Michael L. Seltzer}

\address{Facebook AI}
\email{jaym@fb.com}

\begin{document}

\maketitle
\begin{abstract}
Often, the storage and computational constraints of embedded devices demand that a single on-device ASR model serve multiple use-cases / domains. In this paper, we propose a \emph{Flexible Transducer} (FlexiT) for on-device automatic speech recognition to flexibly deal with multiple use-cases / domains with different accuracy and latency requirements. Specifically, using a single compact model, FlexiT provides a fast response for \emph{voice commands}, and accurate transcription but with more latency for \emph{dictation}. In order to achieve flexible and better accuracy and latency trade-offs, the following techniques are used. Firstly, we propose using domain-specific altering of segment size for Emformer encoder that enables FlexiT to achieve flexible decoding. Secondly, we use Alignment Restricted RNNT loss to achieve flexible fine-grained control on token emission latency for different domains. Finally, we add a domain indicator vector as an additional input to the FlexiT model.
Using the combination of techniques, we show that a single model can be used to improve WERs and real time factor for dictation scenarios while maintaining optimal latency for voice commands use-cases.

\end{abstract}

\noindent\textbf{Index Terms}: Speech recognition, RNN-T, Transformers

\section{Introduction}
On-device automatic speech recognition (ASR) models have been enabled on many embedded devices, including mobile phones, smart speakers, and watches~\cite{he2019rnnt, kim2020attention, venkatesh2021memoryefficient}. On one hand, on-device ASR eliminates the need to transfer audio and recognition results between devices and a server, thus enabling fast, reliable, and privacy-preserving speech recognition experiences. On the other hand, these devices operate with significant hardware constraints: e.g., memory, disk space, and battery consumption. Moreover, the embedded ASR models often serve multiple applications: e.g., video transcription, dictation, and voice commands. Each of these applications has its latency and accuracy requirements. For example, voice assistants demand an ASR model with low latency to respond to user queries as fast as possible. While server-based ASR might rely on running different models for different applications -- more compact models for voice assistants and big, semi-streaming models~\cite{narayanan2020cascaded, cfyeh_asru_2020} for dictation-- the on-device environment prohibits such practice. Due to hardware constraints, and varied requirements of different applications optimizing the model size, compute and accuracy of one single ASR model becomes challenging. In this paper, we take a close look at the scenario where device constrained ASR model needs to be optimized for two different use-cases.
 
The first use case is voice commands, where the latency requirement is strict. Users expect immediate device responses when they ask the speech assistant to turn on the lights or play a song. The second application uses the ASR model for dictation or audio transcription, where accuracy is more important than the model's latency. Recurrent Neural Network Transducer (RNN-T) framework~\cite{he2019rnnt, Graves2012} is widely adopted to provide streaming ASR transcriptions for both voice commands~\cite{shangguan2019optimizing,guo2020efficient} and dictation applications~\cite{shangguan2020analyzing,sainath2020streaming}. 

We focus on Emformer model \cite{emformer} as an audio encoder for RNN-T, which uses both contextual audio information (in the form of an audio chunk) and future audio context (in the form of model look-ahead). A larger model look-ahead permits the model to access more future context and optimize ASR accuracy but hurts model latency. 

This paper proposes a Flexi-Transducer (FlexiT) model that answers the requirements of two streaming ASR use-cases while still staying as one compact model. Further we also show that larger look-ahead enables improve the compute / real time factor trade-offs which help battery consumption.

\vspace{-2mm}
\section{Related Work}
\vspace{-1mm}
Inspired by the successful application of transformer~\cite{Vaswani2017}, many works in ASR also adopted transformer across different model paradigms, such as the hybrid systems~\cite{povey2018time,yongqiang_2019_icassp,frank_zhang_2020,yongqiang2021_icassp}, the encoder-decoder with attention~\cite{dong2018speech,karita2019comparative,sperber2018self,zhou2018syllable,chengyi_wang_2020} and the sequence transducers~\cite{zhang_2020,Yeh_2019,cfyeh_asru_2020}. In this work, we follow the neural transducer paradigm using Emformer~\cite{emformer,yongqiang2021_icassp} and alignment restricted transducer loss~\cite{jay_2020_arrnnt}.

Many ASR applications demand real-time low latency streaming. The block processing method~\cite{dong2019self,emformer,yongqiang2021_icassp} with attention mask modifies the transformer to support streaming applications. In the block processing, self-attention's receptive field consists of one fixed-size chunk of speech utterance and its historical context and a short window of future context. However, the fixed chunk size limits the encoder's flexibility to trade-off latency, real-time factor, and accuracy.

A unified framework is proposed in~\cite{yu2021dualmode} to train one single ASR model for both streaming and non-streaming speech recognition applications. In ~\cite{narayanan2020cascaded} cascaded encoders are used to build a single ASR model that operates in streaming and non-streaming mode. These approaches support one fixed latency for the streaming use-case, and the other use-case is strictly non-streaming. In this paper, we tackle scenarios to support two different streaming ASR use-cases with different latency constraints.

More flexible latency is achieved by~\cite{huang_2020_ddt,tripathi_2020_ymodel,binbin_2020_wenet} where, in the training phase, the model is exposed to audio context variants up to the whole utterance length. In~\cite{huang_2020_ddt}, the authors show that asynchronous revision during inference with convolutional encoders can be used to achieve dynamic latency ASR. The context size selection proposed in these works is purely random during training. In this work, besides random selection, we also explore context size selection based on a priori knowledge about the targeted use cases (domains). The use of domain information to improve the ASR performance of a single model being used to serve different dialects, accents, or use-cases has been studied previously in \cite{li2017multidialect, zhou2018multilingual, sainath2020streaming}

For RNN-T models, both the encoder's context size and the potential delay of token emissions contribute to model latency. It is well known that streaming RNN-T models tend to emit ASR tokens with delay. Techniques like Ar-RNN-T \cite{jay_2020_arrnnt}, Fast-emit \cite{yu2021fastemit}, constrained alignment approach \cite{wordtimings, sak2015fast} or late alignment penalties~\cite{li2020fast} are used to mitigate token delays. This work further extends the alignment restricted transducer~\cite{jay_2020_arrnnt} with task-dependent right buffer restriction to control the token emission latency for different use-cases. 

\section{Methodology}
In this section, we outline the three proposed techniques for FlexiT. Note that we focus on demonstrating these techniques on an Emformer~\cite{emformer}, but the techniques could be extended to other encoder architectures that support dynamic segment size selection.

\vspace{-3mm}
\begin{figure}[h]
    \includegraphics[width=0.4\textwidth]{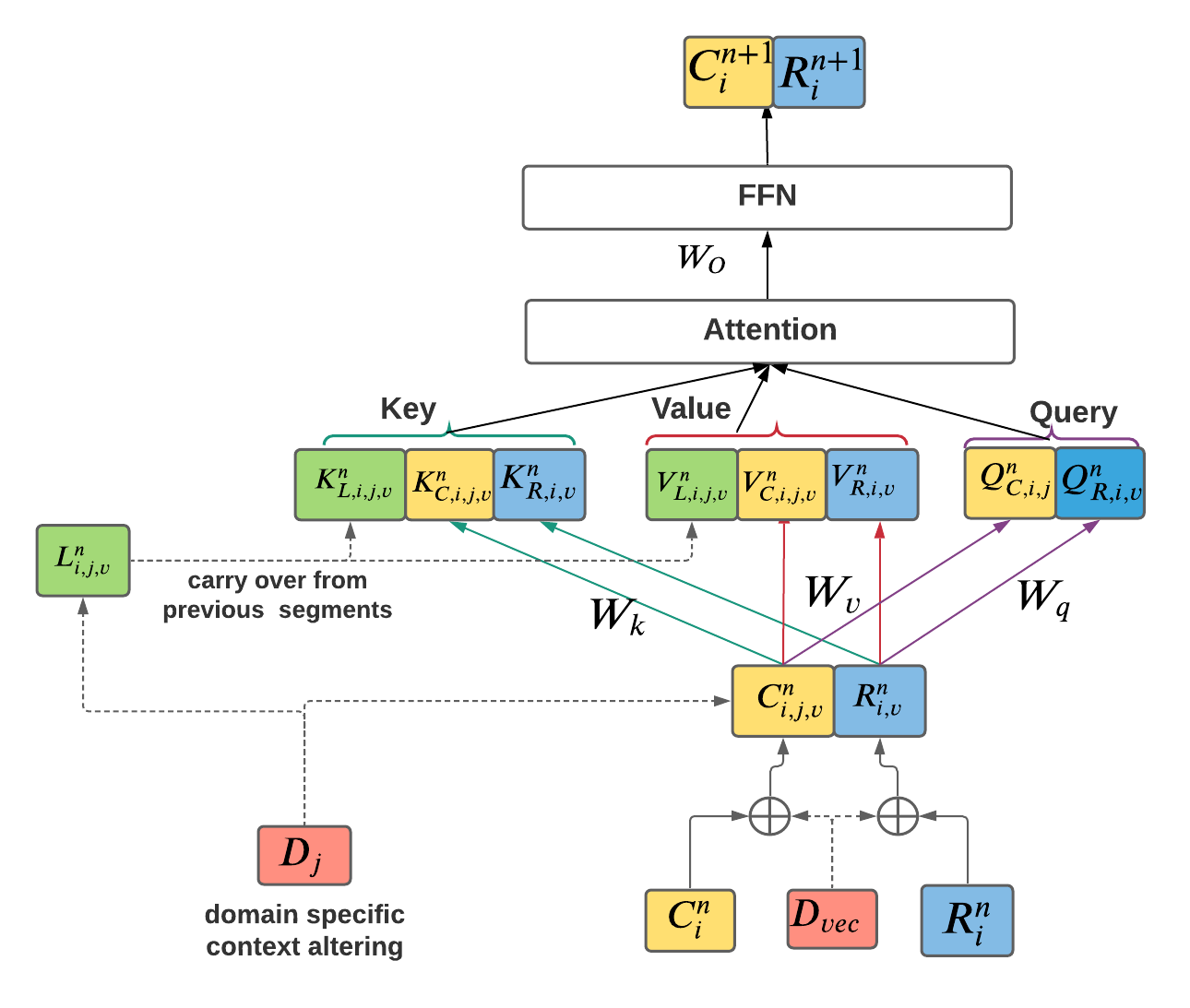}
    \caption{Illustration for domain-specific context size and the injection of domain vectors in an Emformer layer}
    \label{fig:Emformer_domain_specific}
    \vspace{-1.7em}
\end{figure}

\subsection{Domain Specific Segments in Emformer Encoder} 
\label{subsec:domain_specific_Emformer_segments}

FlexiT is shown in Figure~\ref{fig:Emformer_domain_specific}, with notations similar to those in~\cite{emformer}. Input to the Emformer layer concatenates a sequence of audio features into segments $C^n_i$, \dots $C_{I-1}^n$, where $i$ is the index of a segment and $n$ the layer's index. The corresponding left and right contextual blocks $L_i^n$ and $R_i^n$ are concatenated together with $C^n_i$ to form contextual segment $X_i^n = [L_i^n, C_i^n, R_i^n]$. In FlexiT, we do not use the memory vector in the original Emformer layers. We propose to dynamically alter contextual block $L_i^n$, $C_i^n$ by selecting the number of segments $I$ dynamically. We explore both random context segment selection, as well domain-dependent context segment selection during training. 

During training, we ensure every input batch consists of utterances from the same domain. Let $D_j$ denote the domain being used for training the current batch. A domain-specific context altering operation is performed such that the vector $X_i^n$ is modified to $[L_{i,j}^n, C_{i,j}^n, R_i^n]$, conditioned on $D_j$. 
More concretely, according to the desired domain specific context size, $C_i^n$ is split into $[C_{iLeft}^n$, $C_{iRight}^n]$. We then set $L_{i,j}^n = [L_i^n, C_{iLeft}^n]$, and $C_{i,j}^n = C_{iRight}^n$.
The domain-specific selection of $C_{i,j}^n$ provides flexible latency for decoding and, at the same time, as suggested by our results in Section~\ref{subsection:combined}, helps to improve the speech recognition model's robustness.

\vspace{-2mm}
\subsection{Adding Domain Vector in Emformer Encoder}
\label{subsec:domain_vector}
To improve the ASR model's capability to learn domain-specific features, we append a domain vector to inputs of each layer of the Emformer encoder.  The domain vector is simply represented by using 1-hot representation~\cite{li2017multidialect}, the value of which depends on whether the training sample comes from $V_{Cmd}$ or Dictation domain. More concretely, as illustrated in Figure \ref{fig:Emformer_domain_specific} let $D_{vec}$ denote the 1-hot domain vector representation. We concatenate $D_{vec}$ to all components of $X_i^n$, to obtain concatenated input vectors $[L_{i,j,v}^n, C_{i,j,v}^n, R_{i,v}^n]$. These are used as inputs to the $n^{th}$ Emformer layer while training.

\vspace{-2mm}
\subsection{Domain Specific Alignment Restrictions Loss}
In~\cite{jay_2020_arrnnt} using pre-computed token level alignment information, configurable thresholds \textit{$b_l$}, \textit{$b_r$} are used to restrict the alignment paths used for RNN-T loss computation during training. Note that the right-buffer \textit{$b_r$} can be made stricter to ensure earlier token emissions, but stricter \textit{$b_r$} also leads to increased WER. In this work, our goal is to optimize the WERs for the dictation domain while maintaining low latency for the $V_{Cmd}$ domain. Therefore, we propose using domain-specific alignment restriction thresholds while optimizing the loss. We analyze domain-specific WER and token emission latency in Section \ref{sec:Results}.

\section{Experimental Setup}
\label{sec:experimental_setup}
\vspace{-2mm}

\subsection{Datasets}
\subsubsection{Training Data}\label{subsubsec:training_dataset}
\vspace{-1mm}

We run our experiments on data-sets that contains ~20K hours of human-transcribed data from 2 different domains. 

\textbf{Voice Commands} ($V_{Cmd}$ ) dataset combines two sources. The first source is in-house, human transcribed data recorded via mobile devices by 20k crowd-sourced workers. The data is anonymized with personally identifiable information (PII) removed. We distort the collected audio using simulated reverberation and add randomly sampled additive background noise extracted from publicly available videos. The second source came from  1.2 million voice commands (1K hours), sampled from production traffic with PII removed, audio anonymized, and morphed. Speed perturbations~\cite{ko2015audio} are applied to this dataset to create two additional training data at 0.9 and 1.1 times the original speed. We applied distortion and additive noise to the speed perturbed data. From the corpus, we randomly sampled 10K hours.

\textbf{Dictation} (open-domain) dataset consists of 13K hours of data sampled from English public videos that are anonymized with PII removed and annotator transcribed. We first apply the same above-mentioned distortions and then randomly sample 10k hours of the resultant data.

\subsubsection{Evaluation Datasets}
For evaluation, we use the following datasets, representing two different domains:

\textbf{Voice Commands} evaluation set consists of 15K hand-transcribed anonymized utterances from volunteer participants as part of an in-house pilot program.   

\textbf{Dictation} evaluation set consists of 66K hand-transcribed anonymized utterances from vendor collected data where speakers were asked to record unscripted open domain dictation or voice conversations.

\subsection{Evaluation Metrics}\label{evalmetrics}
To measure the model's performance and analyze trade-offs, we track the following metrics: 

\textbf{Accuracy:} We use word-error-rate (WER) to measure model accuracy on evaluation sets. Note that we measure the WERs for dictation domain without end-pointer and the $V_{Cmd}$ domain with end-pointer. We also keep track of the deletion errors (DEL), which are proportional to early cutoffs in $V_{Cmd}$. 

\textbf{Latency:} We measure model latency on $V_{Cmd}$ domain using following metrics: 
\begin{enumerate}[leftmargin=0cm,itemindent=.5cm,labelwidth=\itemindent,labelsep=0cm,align=left]
\vspace{-1mm}
\item \emph{Token Finalization Delay (FD):} as defined in \cite{jay_2020_arrnnt} is the audio duration between the time when user finished speaking the ASR token, and the time when the ASR token was surfaced as part of 1-best partial hypothesis, also referred as emission latency in~\cite{yu2021dualmode}, or user-perceived latency in ~\cite{pratap2020scaling}. 
\vspace{-1mm}
\item \emph{Endpointing Latency (L):} \cite{shanon46162, jay_2020_arrnnt} is defined as the audio time difference between the time end-pointer makes end-pointing decision and the time user stops speaking.
\end{enumerate}

We track the Average Token Finalization Delay and Average Endpointing latency ($L_{Avg}$) metrics. In all experiments, we use a fixed neural end-pointer (NEP) \cite{shanon46162}, running in parallel to ASR being evaluated every 60ms to measure $V_{Cmd}$ domain metrics. A detailed study with other end-pointing techniques besides NEP (static, E2E \cite{joineendpoint}) is beyond the scope of this paper. 

\textbf{ASR Compute:} On device power consumption / battery usage is usually well co-related with the amount of compute being used by the ASR model. We use the real time factor (RTF) measured on a real android device as an indirect indicator of the of the model's compute usage. 

\subsection{Experiments}
We use the RNN-T model with Emformer encoder \cite{yangyangshi_interspeech_2020}, LSTM with layer norm as predictor, and a joiner with 45M total model parameters. As inputs, we use 80-dim log Mel filter bank features at a 10 ms frame rate. We also apply SpecAugment [28] without time warping to stabilize the training. We use a stride of 6 and stack 6 continuous vectors to form a 480 dim vector projected to a 512 dim vector using a linear layer. The model has 10 Emformer layers, each with eight self-attention heads and 512 dimension output. The inner-layer has a 2048 dimension FFN with a dropout of 0.1.  We use Alignment Restricted RNN-T loss \cite{jay_2020_arrnnt} using a fixed left-buffer $b_{l}$ of 300ms, while varying right-buffer $b_{r}$ parameter. All models are trained for 45 epochs using a tri-stage LR scheduler with ADAM optimizer and base LR of 0.005. 

We study the best way to integrate domain information into FlexiT by experimenting with how domain-conditioned Ar-RNN-T buffer sizes, Emformer context sizes (\emph{Emf Ctx}), the domain one-hot vector input, and their combined interactions work to improve the model's performance in $V_{Cmd}$ and dictation domains.

\begin{enumerate}[leftmargin=0cm,itemindent=.5cm,labelwidth=\itemindent,labelsep=0cm,align=left]
\vspace{-1mm}
    \item \textbf{Fixed Emformer Context:}
    We first fix the \emph{Emf Ctx} per experiment and analyze how Ar-RNN-T right buffer size $r_b$ and Domain Vector impact model performances.
    \begin{enumerate}[leftmargin=0.2cm,itemindent=.5cm,labelwidth=\itemindent,labelsep=0cm,align=left]
    \item \emph{Fixed Ar-RNN-T without Domain Vector:} As baselines ($B_1$-$B_3$), we train models using 120, 300 and 600ms \emph{Emf Ctx}. We sweep the range of Ar-RNN-T $b_{r}$ with (120, 300, 420) ms, (300, 600, 900) ms, (600, 900, 1200) ms respectively.
    \item \emph{ Domain Ar-RNN-T without Domain Vector:} In this experiment, we study if domain specific Ar-RNN-T $b_r$ sizes help to improve the WER-latency trade-off. We train models $C_2$,$C_3$ similar to $B_2$,$B_3$ but also adding domain specific Ar-RNN-T $b_r$ of (120, 600)  while for $B_3$ we use $b_r$ of (120, 900) for voice commands / dictation domains. 
    \item \emph{ Fixed Ar-RNN-T with Domain Vector:} We also train models ($D_1-D_3$) where we concatenate domain vector to Emformer layer inputs. The models are trained using 120, 300 and 600ms \emph{Emf Ctx} and Ar-RNN-T $b_{r}$ of 420ms, 600ms and 900ms respectively.
     \item  \emph{ Domain Ar-RNN-T with Domain Vector:} To analyze if domain vector enables the model to learn to emit tokens with different latency, we also train a models $E_2$,$E_3$ using similar configuration as $D_2$,$D_3$ but also adding domain specific Ar-RNN-T threshold $b_r$ of (120, 600) for $E_2$ (120, 900) for $E_3$.
    \end{enumerate}
    \item \textbf{Random / Domain Specific Emformer Context:}  We analyze randomly selected \emph{Emf Ctx} during training as described in \ref{subsec:domain_specific_Emformer_segments}. Context size is randomly selected from 120ms to 1200ms. As shown later in Section \ref{sec:Results} we find domain specific Ar-RNN-T $b_r$ of 420, 900ms achieves best results. Therefore, to reduce the number of combinations in this experiment, we fix 
    $b_{r}$ of 420, 900ms and run experiments $R_1$ and $R_2$ without and with use of Domain vector respectively. Lastly, we analyze using domain specific \emph{Emf Ctx} during training using domain-specific Ar-RNN-T $b_{r}$ of 420, 900ms. Corresponding experiments $S_1$ and $S_2$ without and with Domain vector respectively.
\end{enumerate}

\textbf{Inference:} We always evaluate models such that for each domain, the training time \emph{Emf Ctx} matches the context provided to the encoder while doing inference. Only exception being Random Emformer Context experiments $R_1$ and $R_2$ where we use inference context size of 120ms, 600ms for $V_{Cmd}$  and dictation domains respectively.
\begin{figure}[h]
    \includegraphics[width=0.45\textwidth]{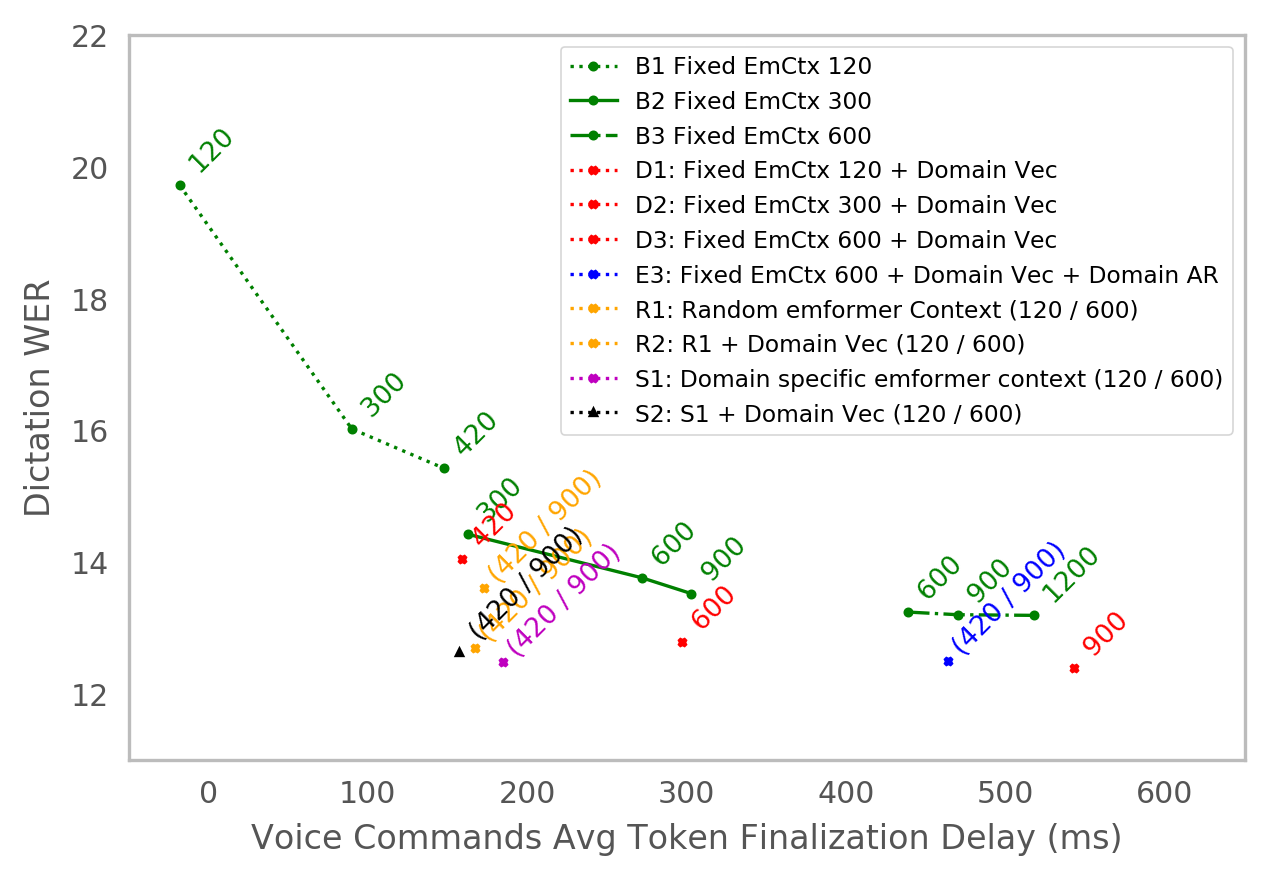}
    \caption{Dictation WERs and $V_{Cmd}$  token finalization delay tradeoffs for various experiments. The labels in the plot show the Ar-RNN-T $b_r$ used for the experiment. Increasing $b_r$ as well as Emformer context improves WER, but degrades latency. Methods ($R_2$, $S_2$) achieve best trade-offs.}
    \label{fig:wer_latency_tradeoff}
\end{figure}

\vspace{-1em}
\section{Results and Analysis}
\label{sec:Results}

\subsection{Emformer Context and Ar-RNN-T Thresholds}

\begin{table}[h]
\centering
\begin{tabular}{|p{0.2cm}| p{0.7 cm} |p{0.9cm}|p{0.6cm} p{0.6cm} p{0.6cm}| p{0.4cm} p{0.5cm} |}\hline
    &\textbf{Emf Ctx} & \textbf{AR $b_r$}  & \textbf{Dict WER} & \textbf{$V_{Cmd}$  WER} & \textbf{$V_{Cmd}$  DEL} & \textbf{Avg FD} & $L_{Avg}$
    \\\hline
    $B_1$ & 120 & 420 & 15.4 & 6.7 & 2.8 & 148 & 449  
    \\$B_2$ & 300 & 600 & 13.8 & 7.4 & 3.6 & 272 & 463  
    \\$B_3$ & 600 & 900 & 13.2 & 9.7 & 5.4 & 470 & 505  
    \\\hline
    $C_2$ & 300 & 420/600 & 13.7 & 7.5 & 3.6 & 271 & 482 
    \\$C_3$ & 600 & 420/900 & 13.2 & 10.8 & 6.4 & 483 & 548  
    \\\hline
\end{tabular}
\vspace{0.5em}
\caption{\textbf{Fixed Emformer Context without Domain Vector}: Dictation WERs improve with larger Emf Ctx. $V_{Cmd}$  WERs and Deletions with neural endpointer, also increase due to increased Avg($FD$).}
\label{table:fixed_emf_no_domain_vec}
\vspace{-2.0em}
\end{table}

Figure \ref{fig:wer_latency_tradeoff} demonstrates the trade-offs as we vary Emformer context and the Ar-RNN-T right-buffer $b_r$. Dictation WER improves as we increase the Emformer context in experiments $B_1$-$B_3$. Similarly, for a fixed Emformer context, WERs improve with larger Ar-RNN-T $b_r$ parameter. Note that only hand-picked variants are detailed in Table \ref{table:fixed_emf_no_domain_vec}.

On the other hand, $V_{Cmd}$ domain's $L_{avg}$ also increases. To achieve low $L_{Avg}$ throughout experiments, which is important for a better user experience, we use a fixed NEP. Delays in token emissions result in more early cuts and increased deletion errors as shown in Table \ref{table:fixed_emf_no_domain_vec},\ref{table:fixed_emf_with_domain_vec}. To achieve best latency and simultaneously reduce early cuts, we must maintain a smaller Emformer context and maintain a strict Ar-RNN-T $b_r$ parameter. Therefore, experiments for random and domain-specific Emformer context are performed with $b_r$ 420 and context 120.

\subsection{Domain Vector and Domain specific Ar-RNN-T}
\label{subsection:domain_vector_and_domain_arrnnt}

\begin{table}[h]
\centering
\begin{tabular}{|p{0.2cm}| p{0.7 cm} |p{0.9cm}|p{0.6cm} p{0.6cm} p{0.6cm}| p{0.4cm} p{0.5cm} |}\hline
    &\textbf{Emf Ctx} & \textbf{AR $b_r$}  & \textbf{Dict WER} & \textbf{$V_{Cmd}$  WER} & \textbf{$V_{Cmd}$  DEL} & \textbf{Avg FD} & \textbf{$L_{Avg}$} 
    \\\hline
    $D_1$ & 120 & 420 & 14.0 & 6.8 & 2.9 & 159 & 441  
    \\$D_2$ & 300 & 600 & 12.8 & 7.7 & 3.8 & 297 & 464
    \\$D_3$ & 600 & 900 & 12.4 & 10.5 & 6.2 & 543 & 509
    \\\hline
    $E_2$ & 300 & 420/600 & 12.8 & 7.23 & 3.28 & 263 & 457 
    \\$E_3$ & 600 & 420/900 & 12.5 & 9.6 & 5.7 & 464 & 476  
    \\\hline
\end{tabular}
\vspace{1em}
\caption{\textbf{Fixed Emformer Context with Domain Vector}: With Domain vector $D_1$-$D_3$ achieve better dictation WERs compared to $B_1$-$B_3$. Further, with Domain specific Ar-RNN-T, $E_2$,$E_3$ achieve better latency compared to $D_2$,$D_3$.}
\label{table:fixed_emf_with_domain_vec}
\vspace{-2.0em}
\end{table}
We observe that providing domain vector in encoder improves the WERs for dictation domain in general for all experiments as shown in Table \ref{table:fixed_emf_with_domain_vec}, which is consistent with previous works \cite{li2017multidialect, sainath2020streaming}. Unfortunately, the WER improvements come in tandem with an increased average FD of $V_{Cmd}$ when the models are trained with domain vector (comparing experiments $B_1$-$B_3$ and $D_1$-$D_3$). We hypothesize that this is because in the absence of stricter AR-RNNT $r_b$ restrictions for the $V_{Cmd}$ domain, the model learns to delay token emissions to improve accuracy.

Ar-RNN-T helps achieve fine grained control on token delays~\cite{jay_2020_arrnnt}. However, in multi-domain setting, comparing $C_2$,$C_3$ to $B_2$,$B_3$, we observe that simply imposing domain specific Ar-RNN-T thresholds does not improve $V_{Cmd}$ FD. The use of domain vector, alongside domain specific Ar-RNN-T thresholds, enables us to achieve a more refined control over $V_{Cmd}$ domain's FD. This is demonstrated in Table \ref{table:fixed_emf_with_domain_vec} where $D_1$-$D_3$ have larger Avg(FD) compared to $B_1$-$B_3$, but models $E_2$-$E_3$ learn to explicitly emit $V_{Cmd}$ domain tokens earlier than $C_2$-$C_3$.

\subsection{Random / Domain Specific Emformer Context}
\label{subsection:combined}
\vspace{-1mm}

\begin{table}[h]
\centering
\begin{tabular}{|p{0.2cm}|  p{1.0 cm} | p{0.6cm}|p{0.6cm} p{0.6cm} p{0.6cm}| p{0.4cm} p{0.5cm} |}\hline
    &\textbf{Emf Ctx}&  \textbf{$D_{vec}$}  & \textbf{Dict WER} & \textbf{$V_{Cmd}$  WER} & \textbf{$V_{Cmd}$  DEL} & \textbf{Avg FD} & \textbf{$L_{Avg}$} 
     \\\hline $R_1$ & \multirow{2}{*}{Random} & No  & 13.6 & 7.2 & 3.1 & 173 & 440  
    \\$R_2$ & & Yes  & 12.7 & 7.1 & 3.1 & 167 & 440  
     \\ \hline
    $S_1$ & \multirow{2}{*}{120/600}  & No  & 12.5 & 7.7 & 3.3 & 185 & 458  
    \\
    $S_2$ &  & Yes  & 12.6 & 7.0 & 2.9 & 157 & 450
    \\ \hline
\end{tabular}
\vspace{1em}
\caption{\textbf{Random and Domain Specific Emformer Context}: Experiments use 420 / 900 Ar-RNN-T $r_b$ parameters and 120 / 600 $EmCtx$ during inference.}
\label{table:results_table}
\vspace{-2em}
\end{table}

Results from random context training, $R_1$ suggests that in the absence of domain information, the model achieves worse trade-offs than $V_{Cmd}$ FD of $B_1$ and Dictation WER of $B3$. In $R_2$ adding domain vector, improves the WER for dictation domain significantly which is consistent with Section \ref{subsection:domain_vector_and_domain_arrnnt}. Overall, $R_2$ achieves better trade-offs for optimizing both use-cases.

Experiment $S_1$, which combines domain-specific \emph{Emf Ctx} and uses domain-specific Ar-RNN-T achieves dictation WERs comparable to $D_3$ while enabling reasonable FD for $V_{Cmd}$ domain, which improves on the results of experiment $R_1$ with random \emph{Emf Ctx}. Therefore, we argue that domain-specific \emph{Emf Ctx}, which enables dynamic attention masking per domain, already helps the model learn more robust domain-specific features, even in the absence of a domain vector. 

Finally, similar to Section \ref{subsection:domain_vector_and_domain_arrnnt} further addition of domain vector in experiment $S_2$ also enables the model to achieve better FD, thus improving the deletion errors from model $S_1$. This achieves the best tradeoffs in dictation domain WER, and $V_{Cmd}$ domain FD and WER with endpointing enabled.

\begin{figure}[h]
    \includegraphics[width=0.48\textwidth]{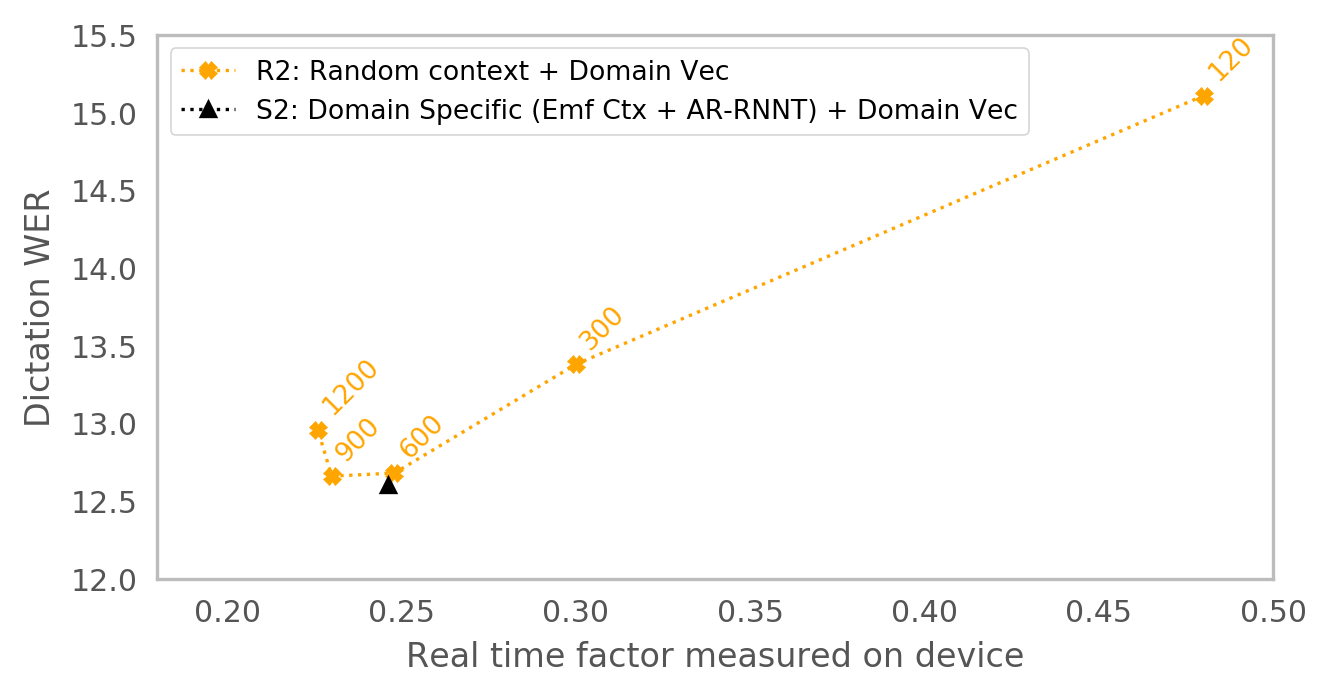}
    \caption{Dictation WERs and Real Time Factor measured on an android device. The labels show chunk size (ms) used during inference, $S_2$ was evaluated with 600ms chunk size.}
    \label{fig:dictation_wer_rtf}
    \vspace{-1em}
\end{figure}
In Figure \ref{fig:dictation_wer_rtf} we analyze the RTF of FlexiT models. For $R_2$ we evaluate the RTF and WERs while varying inference context size. We observe that the RTF reduces with larger context size, which is mainly because of improved batching inside Emformer layers across time dimension. Better RTFs typically are co-related with better compute and power consumption.

\section{Conclusion}
This paper proposed a single \emph{Flexi-Transducer} (FlexiT) model that supports domain-dependent trade-off of latency and accuracy. The domain-specific or random context modeling is achieved jointly via segment size altering operation for encoder and the domain vector. Ar-RNN-T loss imposes a domain-specific constraint to limit the token emission latency for different domains. Using the combination of techniques, we achieve better WER, RTF and latency trade-offs when a single model supports multiple streaming ASR use-cases.

\vspace{0.8em}
\textbf{Acknowledgment:}
\label{sec:acknowledgment}
We'd like to thank Rohit Prabhavalkar and Jiatong Zhou for valuable discussions and advice.

\bibliographystyle{IEEEtran}

\bibliography{mybib}

\end{document}